\documentclass[a4paper,oneside,12pt,english]{article}

\usepackage{epsfig,array,amsmath,amssymb,psfrag,graphicx}

\newcommand{\beq}{\begin{equation}}
\newcommand{\eeq}{\end{equation}}
\newcommand{\bea}{\begin{eqnarray}}
\newcommand{\eea}{\end{eqnarray}}

\newcommand{\RR}{\mathbb{R}}

\newcommand{\mc}[1]{\mathcal{#1}}
\newcommand{\mb}[1]{\mathbf{#1}}

\newcommand{\CQG}{{\it Class. Quant. Grav. }}
\newcommand{\PRD}{{\it Phys. Rev.} D }

\begin{document}

\title{Test of the weak cosmic censorship conjecture with a charged scalar field and dyonic Kerr-Newman black holes}

\author{\small 
G\'abor 
Zsolt T\'oth\thanks{email: tgzs@rmki.kfki.hu}
\\[4mm] 
\small Institute for Particle and Nuclear Physics \\
\small Wigner Research Centre for Physics, Hungarian Academy of Sciences \\
\small Konkoly Thege Mikl\'os \'ut 29-33\\
\small H-1121 Budapest, Hungary\\
\date{}
}
\maketitle

\begin{abstract}
A thought experiment considered recently in the literature, in which it is investigated whether a
dyonic Kerr-Newman black hole can be destroyed by overcharging or overspinning it past extremality 
by a massive complex scalar test field, is revisited.  
Another derivation of the result that this is not possible, i.e.\ the weak cosmic censorship is not violated in this thought 
experiment, is given.   
The derivation is based on conservation laws, on a null energy condition, 
and on specific properties of the 
metric and the electromagnetic field of
dyonic Kerr-Newman black holes. The metric is kept fixed, whereas the
dynamics of the electromagnetic field is taken into account. 
A detailed knowledge of the solutions of the equations of motion 
is not needed.   
The approximation in which the electromagnetic field is fixed is also considered, 
and a derivation for this case is also given. 
In addition, an older version of the thought 
experiment, in which a pointlike test particle is used, is revisited.
The same result, namely the non-violation of the cosmic censorship, is rederived in 
a way which is simpler than in earlier works.
\end{abstract}

\vfill\eject

\section{Introduction}
\label{sec.intr}

If a spacetime contains a singularity 
not hidden behind an event horizon (a ``naked" singularity), then 
far away observers can receive signals coming from this singularity. However, 
initial conditions cannot be specified at a singularity, therefore a singularity that is not behind an event horizon
prevents predictability in the spacetime that contains it. For this reason, it is conjectured that  naked singularities 
cannot be produced in a physical process from regular initial conditions, if the matter involved in the process has reasonable properties.  
This conjecture, first stated by Penrose \cite{Penrose}, is known as the weak cosmic censorship conjecture (WCCC) 
(for a textbook exposition, see e.g.\  section 12.1 
of \cite{WaldGR}), and it is one of the major unsolved problems of classical general relativity to decide whether it is correct.

In the absence of a general proof, the validity of the WCCC has been checked in several special cases by studying the evolution of initially 
regular physical systems. A possible test is to throw a small particle at a Kerr-Newman black hole and to see if
an overextremal Kerr-Newman spacetime, which contains a naked singularity, 
can arise after the particle has been absorbed by the black hole.
Wald \cite{Wald2} was the first who considered this thought experiment, and he showed that an 
extremal Kerr-Newman black hole cannot be overcharged or overspin by throwing a 
pointlike test particle with electric charge into it. A simpler derivation of this result was given by Needham \cite{Needham}. 
Hiscock \cite{Hiscock} and
Semiz \cite{Semiz2} extended Wald's result to the case of dyonic Kerr-Newman black holes, 
which are rotating black holes with both electric and magnetic charge. The derivations presented in \cite{Hiscock,Semiz2} are generalizations of the 
derivation in \cite{Wald2}.
Recently Semiz \cite{Semiz1} also studied the case when a complex scalar test field is used instead of a test particle, and found that the WCCC is not 
violated. 
Other results supporting the WCCC were obtained by several authors, 
and there are also claims that the WCCC can be violated, for example by starting from a slightly sub-extremal black hole and 
``jumping over" the extremal case \cite{CG}-\cite{SS}. 
The signatures of naked singularities for the observational verification 
of their existence was also investigated, 
e.g.\ in \cite{VE,VK}.
Reviews on the status of the cosmic censorship conjecture can be found in \cite{Wald0}-\cite{Krolak}.

A dyonic Kerr-Newman black hole can be characterized by four parameters, which are the mass $M$, the angular momentum per unit mass
$a$, the electric charge $Q_e$ and the magnetic charge $Q_m$. The angular momentum of the black hole is $J=aM$, and
$Q_m=0$ corresponds to a usual Kerr-Newman black hole. 
The metric of the dyonic Kerr-Newman black hole spacetime with parameters $(M,a,Q_e,Q_m)$
is the same as the Kerr-Newman metric with parameters $(M,a, e)$, $e^2 = Q_e^2+Q_m^2$, where $e$ denotes the electric charge parameter 
of the Kerr-Newman metric.
The parameters have to satisfy the inequality 
\beq
\label{eq.eta}
\eta=M^2-Q_e^2-Q_m^2-a^2\ge 0,
\eeq
otherwise the spacetime contains a naked singularity. The black hole is called extremal if 
$\eta=0$. 
Under certain conditions, the dyonic Kerr-Newman black holes are the only static and asymptotically flat black hole solutions
of the Einstein-Maxwell equations \cite{Mazur,Bunting}.

Under a small change $(dM,dJ,dQ_e,dQ_m)$ of the parameters of the black hole, the change of $\eta$ is 
\beq
\label{eq.deta}
d\eta=2\frac{M^2+a^2}{M}\left( dM-\frac{a}{M^2+a^2}dJ-\frac{Q_e M}{M^2+a^2}dQ_e-\frac{Q_m M}{M^2+a^2}dQ_m  \right).
\eeq
In the thought experiments discussed in \cite{Wald2,Needham,Hiscock,Semiz2,Semiz1} it is assumed that initially 
one has an extremal dyonic Kerr-Newman black hole, 
which then absorbs a small amount of matter, and finally settles down in another 
dyonic Kerr-Newman state with slightly different parameters. 
If one calculates the change $(dM,dJ,dQ_e,dQ_m)$ of the parameters in this process, one should find $d\eta\ge 0$; a result $d\eta < 0$ would indicate
a possible violation of the WCCC. 
In \cite{Wald2,Needham,Hiscock,Semiz2,Semiz1} the change
$(dM,dJ,dQ_e,dQ_m)$ of the parameters were calculated in the approximation that the metric 
is fixed during the process and, as mentioned above, the result $d\eta\ge 0$ was found.

In \cite{Wald2,Needham,Hiscock,Semiz2}, 
where the thought experiment with a pointlike test particle is considered,
not only the metric but also the electromagnetic field is taken to be fixed 
in the calculation of $(dM,dJ,dQ_e,dQ_m)$.
In \cite{Semiz1}, however, where the case of the test field is considered, 
the electromagnetic field is also taken to be dynamical, although with the restriction that
free electromagnetic radiation that is not tied to the electric current is not present.

In this paper we revisit the thought experiment studied by Semiz \cite{Semiz1}, 
in which it is investigated whether a
dyonic Kerr-Newman black hole can be destroyed by overcharging or overspinning it past extremality by a massive complex scalar test field. 
We give a different and simpler derivation of the result that $d\eta \ge 0$ in the extremal case, 
which indicates that the
black hole remains intact and thus the WCCC is not  violated. 
Our derivation
is based on conservation laws, 
on a null energy condition, 
and on certain specific properties of the 
metric and the electromagnetic field of
dyonic Kerr-Newman black holes.  
In contrast with the derivation in \cite{Semiz1}, 
in our derivation it is not necessary to have a 
detailed knowledge of the solutions of the  equations of motion, and we
do not 
impose the restriction on the electromagnetic field that is imposed in \cite{Semiz1} either.  

We also consider the older version of the thought 
experiment in which a pointlike test particle is applied, and we present a slightly different and simpler derivation than 
those in \cite{Hiscock,Semiz2} for this case as well. This is an extension of Needham's derivation \cite{Needham} for Kerr-Newman black holes 
to the dyonic case. 

In these derivations we make the same 
approximations as are made in \cite{Wald2,Needham,Hiscock,Semiz2,Semiz1}, 
i.e.\ in the case of the pointlike test particle 
both the metric and the electromagnetic field is kept fixed, 
whereas in the case of the test field only the metric is kept fixed. 
Nevertheless, one can also consider the scalar test field in fixed gravitational and electromagnetic fields,
therefore we present a derivation for this case as well.  
Apart from certain differences, this derivation is similar to the one for the case of non-fixed 
electromagnetic field.

In our arguments we do not restrict ourselves to extremal black holes, 
but rather we derive an inequality in each case that is valid 
for any values of the black hole parameters and that gives the desired result in the extremal case.

It is worth noting that processes similar to those mentioned above 
are dealt with 
in the derivation of ``physical process versions" of the first law of black hole mechanics \cite{Wald3}-\cite{Nielsen}.
These derivations are similar to some extent to those that we present for the case of the scalar field, 
nevertheless the dynamics of the metric has an important role in them.

The paper is organized as follows. 
In section \ref{sec.knd} further necessary information about the dyonic Kerr-Newman black holes is recalled.
In section \ref{sec.pp} the thought experiment with pointlike particle is discussed and the derivation of the result 
that the cosmic censorship is not violated is given.
In section \ref{sec.ft} the version of the thought experiment in which a scalar field is used is considered, and our derivations of the 
non-violation of the cosmic censorship for this case are presented.
In the Appendix Noether's theorem, as it is used in section \ref{sec.ft}, is described.

\section{The dyonic Kerr-Newman black holes}
\label{sec.knd}

The metric of the dyonic Kerr-Newman black hole spacetime with parameters\\ 
$(M,a,Q_e,Q_m)$ can be given in a standard form as
\begin{eqnarray}
&& ds^2=g_{\mu\nu}\, dx^\mu dx^\nu = \nonumber \\
&&\qquad -\left( \frac{\Delta-a^2\sin^2\theta}{\Sigma} \right) dt^2 -\frac{2a\sin^2\theta (r^2+a^2-\Delta)}{\Sigma}\, dt d\phi \nonumber \\
&&\qquad +\frac{\Sigma}{\Delta}\, dr^2
+ \Sigma\, d\theta^2 + \left( \frac{(r^2+a^2)^2-\Delta a^2\sin^2\theta}{\Sigma} \right)  \sin^2\theta \,  d\phi^2,
\end{eqnarray}
where
\begin{eqnarray}
\Sigma & = & r^2 + a^2 \cos^2\theta\\
\Delta & = & r^2 + a^2 + Q_e^2 +Q_m^2 -2Mr.
\end{eqnarray}
The signature of this metric is $(-+++)$.

The electromagnetic field of a dyonic Kerr-Newman black hole has the vector potential 
\beq
A=Q_e A_e+ Q_m A_m\ , 
\eeq 
where 
\beq
\label{eq.ae}
A_e=-\frac{r}{\Sigma}dt  +  \frac{a r \sin^2\theta}{\Sigma}d\phi
\eeq
and
\beq
\label{eq.am}
A_m=\frac{ a\cos\theta}{\Sigma}dt  +
\left[\tilde{C}-  \cos\theta \frac{r^2+a^2}{\Sigma} \right]   d\phi.
\eeq
The electromagnetic field derived from $A_m$ is dual to the electromagnetic field derived from $A_e$.
The electromagnetic field does not depend on the constant $\tilde{C}$,
which can be used, by setting $\tilde{C}=1$ or $\tilde{C}=-1$, to eliminate 
the Dirac string singularity of $A_m$ along the positive or negative $z$ axis ($\theta=0$ and $\theta=\pi$), respectively. We set $\tilde{C}$ to zero for a reason that is explained below.

In the following sections various quantities will be considered at the future event horizon. 
Since the Boyer-Lindquist coordinates $(t,r,\theta,\phi)$ do not cover the future event horizon,   
Eddington--Finkelstein-type ingoing horizon-penetrating coordinates, 
denoted by $(\tau,r,\theta,\varphi)$, will be used. 
These coordinates can be introduced by the transformation
\begin{eqnarray}
\label{tr1}
\tau & = & t-r+\int dr\, \frac{r^2+a^2}{\Delta} \\
\label{tr2}
\varphi & = & \phi+ \int dr\, \frac{a}{\Delta} \,.
\end{eqnarray}
The future event horizon is located in these coordinates at the constant value 
\beq
r_+=M+\sqrt{M^2-(a^2+Q_e^2+Q_m^2)}
\eeq 
of $r$, and the metric is non-singular in these points. 
In the extremal case  
\beq
r_+=M.
\eeq
The $(\tau+r, \theta,\varphi) = constant$ lines are ingoing null geodesics, and 
there exists an $r_0< r_+$ such that 
the $\tau=constant$ hypersurfaces
are spacelike in the domain $r_0< r$.

The $r$ component $(A_e)_r$ of $A_e$ with respect to the coordinates $(\tau,r,\theta,\varphi)$
is singular at the event horizon, but this singularity can be eliminated by 
the 
gauge transformation $A_e \to A_e - \frac{r}{\Delta} dr$. 
After this gauge transformation 
\beq
\label{eq.ae2}
A_e=-\frac{r}{\Sigma}d\tau  +  \frac{a r \sin^2\theta}{\Sigma}d\varphi
-\frac{r}{\Sigma} dr.
\eeq

The $r$ component of $A_m$ with respect to the coordinates $(\tau,r,\theta,\varphi)$ is also singular 
if $\tilde{C}\ne 0$, therefore we set $\tilde{C} = 0$. Nevertheless, in order to treat the Dirac string singularity of $A_m$, we introduce an explicit gauge parameter into it by adding $Cd\varphi$, where $C$ is a real constant. Thus  
\beq
\label{eq.am2}
A_m=\frac{ a\cos\theta}{\Sigma}d\tau  +
\left[C-  \cos\theta \frac{r^2+a^2}{\Sigma} \right]   d\varphi
+\frac{a\cos\theta}{\Sigma}dr.
\eeq
$A_m$ has a string singularity along the $z$ axis (which corresponds to $\theta=0$ and $\theta=\pi$) 
because $d\varphi$ is singular here, and its coefficient $(A_m)_\varphi$
does not cancel this singularity. However, in the special cases $C=1$ and $C=-1$ the singularity is cancelled
along the positive $z$ axis ($\theta=0$) or along the negative $z$ axis ($\theta=\pi$), respectively.
The string singularity can therefore be avoided by using two domains that cover the whole spacetime region
of interest 
in such a way that one of the domains contains the entire positive $z$ axis but is well separated from 
the negative $z$ axis and the other one contains the entire negative $z$ axis but is separated from the 
positive $z$ axis. In the first domain the $C=1$ gauge is used then, and in the second domain the $C=-1$ gauge. 
Suitable domains are given by $r_0<r$, $0\leq \theta \leq \pi/2$ and $r_0<r$, $\pi/2 < \theta \leq \pi$, for example. 
These domains will be denoted by $\mc{D}_+$ and $\mc{D}_-$.
The transition between the two domains involves a gauge transformation, which has to be kept in mind
in particular calculations.
This approach to treating the string singularity of $A_m$ 
was proposed in \cite{WY} and was also taken in \cite{Semiz2,Semiz1}.

In the rest of the paper we use only the coordinates $(\tau,r,\theta,\varphi)$, 
and we also use the notation $\omega$ for the one-form
$dr$ (the exterior derivative of the coordinate function $r$). 
$A_e$, $A_m$ and $A$ will denote (\ref{eq.ae2}), (\ref{eq.am2}) and $A=Q_e A_e+ Q_m A_m$, respectively.

$\partial/\partial \tau$ and $\partial/\partial \varphi$ are Killing fields; 
$\partial/\partial \tau$ is the generator of time translations and $\partial/\partial \varphi$ is the generator of rotations around the 
axis of the black hole.
$A_e$ and $A_m$ are also invariant under 
these symmetries.
The Killing field
\beq
\label{eq.chi}
\chi=\frac{\partial}{\partial \tau} +\Omega_H \frac{\partial}{\partial \varphi}
\eeq
is null at the event horizon,
with  
\beq
\label{eq.omega}
\Omega_H=\frac{a}{r_+^2+a^2}\ ,
\eeq
which is called the angular velocity of the event horizon.
At the event horizon we also have
\beq
\label{eq.a}
(A_e)_\mu\chi^\mu=\frac{-r_+}{r_+^2+a^2}\ ,\qquad
(A_m)_\mu\chi^\mu=C\Omega_H\ ,
\eeq
and  $\omega^\mu$ is parallel to $\chi^\mu$. The relation between $\omega^\mu$ 
and $\chi^\mu$ at the event horizon is
\beq
\label{eq.rho}
\omega^\mu=\frac{r_+^2+a^2}{r_+^2+a^2\cos^2\theta} \chi^\mu\ ,
\eeq
thus $\omega^\mu$ is also future directed.
We introduce the quantity $\Phi_H$ as
\beq
\label{eq.h}
\Phi_H=\frac{r_+ Q_e}{r_+^2+a^2}\ .
\eeq
In the case of Kerr-Newman black holes, $\Phi_H$ is known as the electrostatic potential of the horizon. 

Both $\mc{D}_+$ and $\mc{D}_-$ are contractible domains, therefore any gauge transformation 
takes the form $A \to A + d\Phi$ on $\mc{D}_+$ or on $\mc{D}_-$, where $\Phi$ denotes a real valued function. 
If $d\Phi$ is invariant under the time translation and rotation symmetries of the spacetime, then 
$\Phi$ takes the form $\Phi_0(r,\theta)+c_1\tau + c_2\varphi$, where $c_1$ and $c_2$ are constants, thus the corresponding gauge transformation changes 
the $\tau$ and $\varphi$ components of $A$ only by adding the constants $c_1$ and $c_2$. 
This shows that if  
$A$ (or $A_e$, $A_m$), understood here to be fixed up to gauge transformations, 
is required to be
invariant under time translation and rotation, then 
the $\tau$ and $\varphi$ components of 
$A$ are determined uniquely up to additive constants. These constants can also 
be fixed by requiring that $A_\tau$ 
should tend to $0$ as $r\to\infty$ and
$A$ should not have a Dirac string singularity.
If the $\tau$ and $\varphi$ components of 
$A$ are fixed, then 
the quantity $A_\mu\chi^\mu$ is also fixed, because it depends only on these components of 
$A$.

\section{Thought experiment with a point particle}
\label{sec.pp}

In this section we consider the thought experiment in which a pointlike test particle is thrown at a dyonic Kerr-Newman black hole. 
The extremality of the black hole is not assumed; we derive an inequality that holds for any values of the black hole parameters,
and that becomes the desired inequality $d\eta > 0$ in the extremal case. The argument that we present is an extension of the argument of \cite{Needham} for Kerr-Newman black 
holes to the more general case of dyonic Kerr-Newman black holes. 

As explained in \cite{Semiz2}, the magnetic charge of the test particle can always be set to zero by a duality rotation,
therefore it can be assumed without loss of generality that the test particle has zero magnetic charge.

The Lagrangian of the test particle with mass $m$, electric charge $q$ and zero magnetic charge is 
\beq
\mc{L}=\frac{1}{2}mg_{\mu\nu} \frac{dx^\mu}{ds}\frac{dx^\nu}{ds} + q A_\mu \frac{dx^\mu}{ds}\ , 
\eeq
and its conserved energy and angular momentum are 
\begin{eqnarray}
\label{eq.y1}
E & = & -mg_{\tau\mu}\frac{dx^\mu}{ds}-qA_\tau   \\
\label{eq.y2}
L & = & +mg_{\varphi \mu}\frac{dx^\mu}{ds}+q(A_\varphi - Q_mC) \ .
\end{eqnarray}
The $-qQ_mC$ term on the right hand side is added to cancel the dependence of $A_\varphi$ on the 
gauge parameter $C$. This is important because $C$ has different values in the two domains
$\mc{D}_+$ and $\mc{D}_-$.

By multiplying  (\ref{eq.y2}) by $\Omega_H$  and then subtracting it from (\ref{eq.y1}), 
and taking into account (\ref{eq.chi}), (\ref{eq.omega}), (\ref{eq.a}) and (\ref{eq.h}), it follows immediately that 
if the particle does cross the event horizon, then 
\beq
\label{eq.xw}
-m\frac{dx^\mu}{ds}\chi_\mu= E-\Omega_H L - \Phi_H q
\eeq
holds at the point where the crossing takes place. At this point  
$\frac{dx^\mu}{ds}\chi_\mu < 0$
also obviously holds, since 
$\frac{dx^\mu}{ds}$ is a timelike future directed vector in the case of a massive particle, and 
$\chi^\mu$ is a future directed null vector at the event horizon.
Thus,
\beq
\label{eq.x0}
E-\Omega_H L - \Phi_H q > 0\ .
\eeq 

We note that 
$\frac{dx^\mu}{ds}\chi_\mu$ is just the $r$ component of $\frac{dx^\mu}{ds}$ 
multiplied by a positive number, as can be seen from (\ref{eq.rho}), 
and the $r$ component of $\frac{dx^\mu}{ds}$ is clearly negative or zero for a 
particle moving inward into the black hole.

The change of the black hole parameters $dM$, $dJ$ and $dQ_e$ in (\ref{eq.deta}) can be identified 
with $E$,  $L$ and $q$, respectively, and $dQ_m=0$. 
The inequality (\ref{eq.x0}) together with the relations $dM=E$, $dJ=L$, $dQ_e=q$ and $dQ_m=0$ imply 
\beq
\label{eq.x}
dM-\Omega_H dJ - \Phi_H dQ_e > 0\ .  
\eeq
In the extremal case $\Omega_H=\frac{a}{M^2+a^2}$ and $\Phi_H=\frac{M Q_e}{M^2+a^2}$, thus in this case 
(\ref{eq.deta}) can be written as
$dM-\Omega_H dJ - \Phi_H dQ_e = \frac{M}{2(M^2+a^2)}d\eta $, therefore in the extremal case
(\ref{eq.x}) gives $d\eta > 0$, 
which indicates that the black hole is not destroyed by the absorption of the test particle, 
i.e.\ no violation of the WCCC occurs.

\section{Thought experiment with test fields}
\label{sec.ft}

In this section we consider a similar thought experiment as in section \ref{sec.pp}, but with test fields 
instead of a pointlike test particle. We discuss two different settings, in sections
\ref{sec.ft1} and \ref{sec.ft2}, respectively.

In section \ref{sec.ft1} we consider the process in which a small amount of electrically charged matter, 
described by a complex scalar field $\psi$,
falls into a dyonic Kerr-Newman black hole.
As in section \ref{sec.pp}, we make the approximation
in the calculation of $dM$, $dJ$ and $dQ_e$ that the metric and the electromagnetic field do not change, 
i.e.\ we take the scalar test field to evolve in the fixed gravitational and electromagnetic 
background fields of the black hole. 

In section \ref{sec.ft2} a similar process is considered, with the difference that 
only the gravitational field is kept fixed, which means that 
the test matter also has an electromagnetic field component and the effect of the scalar field on the electromagnetic field is taken into account. 
This is the setting that was considered in \cite{Semiz1}, although in \cite{Semiz1} the
difference of the total electromagnetic field and the electromagnetic field of the black hole
is tied to the electric current (see \cite{Semiz1} for details). 
We do not impose such a restriction on the electromagnetic field.

Our reason for considering also the first case, in which the electromagnetic field is fixed, is 
that this is the one that is obtained from the thought experiment described in section \ref{sec.pp}
if the pointlike test particle
is replaced by a scalar test field in a straightforward manner,
and it also has 
technical differences from 
the second case.

One of the technical differences between the two 
settings is 
that 
the Einstein-Hilbert energy-momentum tensor is conserved and can be used 
to obtain the conserved energy and angular momentum currents only in the
second setting. 
For this reason in the first setting
we use Noether's theorem to find the conserved currents.
Noether's theorem can be used in the second setting as well; 
the currents obtained in this way differ only in divergence terms from 
the currents obtained from the Einstein-Hilbert energy-momentum tensor, and these terms do not 
give any contribution to $dM$ and $dJ$.

In the same way as in section \ref{sec.pp}, the extremality of the black hole is generally not assumed in 
sections \ref{sec.ft1} and \ref{sec.ft2}.

\subsection{Scalar test field}
\label{sec.ft1}

The action of the scalar field in fixed dyonic Kerr-Newman gravitational and electromagnetic fields is 
$\mc{S}=\int \sqrt{- g}\ \mc{L}\ d\tau dr d\theta d\varphi$, with the Lagrangian density
\beq
\label{eq.lagr}
\mc{L}=-g^{\mu\nu}(\partial_\mu-ieA_\mu)\psi^*(\partial_\nu+ieA_\nu)\psi-m^2\psi^*\psi,
\eeq
where $A_\mu$ is the vector potential of the electromagnetic field of 
the black hole as given in section \ref{sec.knd}.

The energy and angular momentum 
Noether currents corresponding to the symmetries generated by $\partial/\partial \tau$ and $\partial/\partial \varphi$ are 
\beq
\sqrt{-g}\, \mc{E}^\mu=\sqrt{-g}\, {T^\mu}_\nu (\partial/\partial \tau)^\nu = \sqrt{-g}\, {T^\mu}_\tau
\eeq
and 
\beq
\sqrt{-g}\, \mc{J}^\mu=\sqrt{-g}\, {T^\mu}_\nu (\partial/\partial \varphi)^\nu = \sqrt{-g}\, {T^\mu}_\varphi \ ,
\eeq 
where  ${T^\mu}_\nu$ is   
\begin{eqnarray}
{T^\mu}_\nu & = &  
 -\frac{\partial\mc{L}}{\partial_\mu\psi}\partial_\nu\psi -\frac{\partial\mc{L}}{\partial_\mu\psi^*}\partial_\nu\psi^*
+{\delta^\mu}_\nu\mc{L} \nonumber \\
 & = &   (\partial^\mu-ieA^\mu)\psi^*\partial_\nu\psi
+(\partial^\mu+ieA^\mu)\psi \partial_\nu\psi^*
+{\delta^\mu}_\nu\mc{L}\ 
\end{eqnarray}
(see Appendix A for more details on Noether's theorem).
The conservation laws for these currents are $\partial_\mu (\sqrt{-g}\mc{E}^\mu)=0$ and $\partial_\mu (\sqrt{-g}\mc{J}^\mu)=0$. ${T^\mu}_\nu$ is introduced only for notational convenience.

The Noether current corresponding to the $\psi\to e^{i\alpha}\psi$, $\alpha\in \RR$ global $U(1)$ symmetry 
is $\sqrt{-g} j^\mu$, where
\beq
j^\mu=ie[\psi^*(\partial^\mu+ieA^\mu)\psi-\psi(\partial^\mu-ieA^\mu)\psi^*]
\eeq
is the electric current. $j^\mu$ also satisfies the equality
$j^\mu=\frac{\partial \mc{L}}{\partial A_\mu}$.

${T^\mu}_\nu$, $\mc{E}^\mu$, $\mc{J}^\mu$ and $j^\mu$
are quantities that transform as proper tensor and vector fields, respectively, under coordinate transformations.
The conservation laws can be written, of course, as $\nabla_\mu  \mc{E}^\mu=0$, 
$\nabla_\mu \mc{J}^\mu=0$ and  $\nabla_\mu j^\mu=0$.

Defining $\hat{T}_{\mu \nu}$ as 
\beq
\label{eq.tmn}
\hat{T}_{\mu\nu}=
(\partial_\mu-ieA_\mu)\psi^* (\partial_\nu +ieA_\nu)\psi
+ (\partial_\mu+ieA_\mu)\psi (\partial_\nu -ieA_\nu)\psi^* + g_{\mu\nu}\mc{L}\ ,
\eeq
we have 
\beq
T_{\mu\nu}=\hat{T}_{\mu\nu} + A_\nu j_\mu\ ,
\eeq
and $\mc{E}^\mu= \hat{T}^\mu_{\ \ \tau} + A_\tau j^\mu$, $\mc{J}^\mu= \hat{T}^\mu_{\ \ \varphi} + A_\varphi j^\mu$.
$\hat{T}_{\mu\nu}$ and $j^\mu$ are gauge invariant and $A_\tau$ does not depend on 
the gauge parameter $C$,
therefore $\mc{E}^\mu$ is also independent of $C$. 
$A_\varphi$ does depend on $C$, however, thus $\mc{J}^\mu$ also depends on it.
For this reason we take the modified definition   
\beq
\label{eq.jmod}
\mc{J}^\mu= \hat{T}^\mu_{\ \ \varphi} + (A_\varphi - Q_m C)j^\mu 
\eeq
for $\mc{J}^\mu$, which eliminates its dependence on $C$. 
The conservation of $\mc{J}^\mu$ is not affected by this modification, 
because $j^\mu$ is conserved.  
The independence of $\mc{E}^\mu$ and $\mc{J}^\mu$ of $C$ is important because 
the value of $C$ is different in the domains $\mc{D}_+$ and $\mc{D}_-$.

The charge flux through the event horizon into the black hole is
\beq
\label{eq.q}
\frac{dQ}{d\tau} = - \int_H \sqrt{-g}\ j^r\, d\theta d \varphi\ ,
\eeq 
where $H$ denotes the two-dimensional surface of the black hole 
(which is the relevant time slice of the event horizon),
and the energy and angular momentum fluxes are 
\begin{eqnarray}
\label{eq.e}
\frac{dE}{d\tau} & = & \phantom{-}\int_H \sqrt{-g}\ \left[{\hat{T}^r }_{\ \, \tau}  + A_\tau j^r \right]   \, d\theta d \varphi\\
\label{eq.l}
\frac{dL}{d\tau} & = & - \int_H \sqrt{-g}\ \left[{\hat{T}^r}_{\ \, \varphi}   +  
(A_\varphi -Q_mC) j^r \right] \, d\theta d \varphi\ ,
\end{eqnarray}
where the quantities in the brackets are $\mc{E}^r$ and $\mc{J}^r$, respectively.
The total energy, angular momentum and charge that falls through the event horizon is 
$\int_{-\infty}^\infty  \frac{dE}{d\tau} d\tau$,   
$\int_{-\infty}^\infty   \frac{dL}{d\tau} d\tau  $
and $\int_{-\infty}^\infty  \frac{dQ}{d\tau} d\tau$, respectively. 

One of the main assumptions of the thought experiment is that the final state
of the physical system is again a dyonic Kerr-Newman state, which means 
that all matter that does not fall through the event horizon is assumed to escape eventually to
infinity. In particular, it is assumed that the energy, angular momentum and charge of the matter
contained in the domain $r_+\le r \le r_m$, given by the integrals 
$-\int_{r_+}^{r_m} dr \int \sqrt{-g}\, \mc{E}^\tau  d\theta d \varphi$, 
$\int_{r_+}^{r_m} dr \int  \sqrt{-g}\, \mc{J}^\tau  d\theta d \varphi$, 
$\int_{r_+}^{r_m} dr \int  \sqrt{-g}\, j^\tau  d\theta d \varphi$, 
go to $0$ as $\tau \to \infty$ for any fixed value of $r_m$.
Under this assumption $dM$, $dJ$ and $dQ_e$ can be identified with 
$\int_{-\infty}^\infty  \frac{dE}{d\tau} d\tau$,   
$\int_{-\infty}^\infty   \frac{dL}{d\tau} d\tau  $
and $\int_{-\infty}^\infty  \frac{dQ}{d\tau} d\tau$,
i.e.\ the change of the mass, angular momentum and 
electric charge of the black hole equals to the  
total energy, angular momentum and electric charge that falls through the event horizon.

From the equations (\ref{eq.q}), (\ref{eq.e}), (\ref{eq.l}) above and from (\ref{eq.chi}), (\ref{eq.omega}), (\ref{eq.a}) and (\ref{eq.h})
it follows immediately that 
\beq
\label{eq.tt}
\int_H \sqrt{-g}\ \hat{T}_{\mu\nu}\omega^\mu \chi^\nu   \, d\theta d \varphi =
\frac{dE}{d\tau}-\Omega_H  \frac{dL}{d\tau}  -  \Phi_H \frac{dQ}{d\tau}\ .
\eeq
Taking into account the relations $dM=\int_{-\infty}^\infty  \frac{dE}{d\tau} d\tau$,   
$dJ=\int_{-\infty}^\infty   \frac{dL}{d\tau} d\tau  $
and $dQ_e=\int_{-\infty}^\infty  \frac{dQ}{d\tau} d\tau$,
\beq
\label{eq.oo}
\int_{-\infty}^\infty d\tau \int_H \sqrt{-g}\ \hat{T}_{\mu\nu}\omega^\mu \chi^\nu   \, d\theta d \varphi =
dM-\Omega_H dJ -  \Phi_H dQ_e
\eeq
is obtained from (\ref{eq.tt}). 
The right hand side in (\ref{eq.oo}) is the same as the left hand side in  
(\ref{eq.x}), and in the extremal case it is $\frac{M}{2(M^2+a^2)}d\eta$. 
Thus the sign of $d\eta$ depends, in the extremal case, on the sign of
$\int_{-\infty}^\infty d\tau \int_H \sqrt{-g}\ \hat{T}_{\mu\nu}\omega^\mu \chi^\nu   \, d\theta d \varphi $.

From the fact that $\chi^\mu$ is a null vector at the event horizon 
and from the form (\ref{eq.tmn}) of $\hat{T}_{\mu\nu}$
it is obvious  
that at the event horizon $\hat{T}_{\mu\nu}$ satisfies the 
inequality $\hat{T}_{\mu\nu}\chi^\mu \chi^\nu \ge 0$  for arbitrary $A_\mu$ and $\psi$ functions.
Taking into consideration (\ref{eq.rho}), this also means that
$\hat{T}_{\mu\nu}\omega^\mu \chi^\nu \ge 0$ holds for the integrand $\hat{T}_{\mu\nu}\omega^\mu \chi^\nu$ on the left hand side of (\ref{eq.oo}), hence 
\beq
\label{eq.y}
dM-\Omega_H dJ -  \Phi_H dQ_e\ge 0\ .
\eeq 
In particular, in the extremal case $d\eta\ge 0$, 
indicating that the WCCC 
is not violated.
The inequality (\ref{eq.y}) is almost identical to (\ref{eq.x}), the only minor difference is that in  (\ref{eq.y}) equality is allowed.

Regarding the condition of strict equality in (\ref{eq.y}), 
$dM-\Omega_H dJ -  \Phi_H dQ_e = 0$ holds if and only if $\chi^\mu(\partial_\mu+ieA_\mu)\psi=0$ everywhere on the future part of the event horizon.
It is easy to see that in this case the charge, energy and angular momentum fluxes
$\frac{dQ}{d\tau}$,  $\frac{dE}{d\tau}$ and $\frac{dL}{d\tau}$ into the black hole are zero, thus $dM=dJ=dQ_e=0$. 
Furthermore, since $\chi^\mu A_\mu $ is a real constant on the event horizon, the equation $\chi^\mu(\partial_\mu+ieA_\mu)\psi=0$ 
implies that either $\psi=0$ everywhere on the 
future part of the event horizon, or $\psi$ is of the form $\psi=\psi_0e^{i\alpha \tau}$, where $\alpha\in\RR$ and $\psi_0\ne 0$,    
along some integral curves of $\chi^\mu$. The second possibility is excluded if we assume $\lim_{\tau\to -\infty} \psi =0$ at the event horizon.

We close this section with two remarks.
Although the expressions $\mc{E}^\mu = \hat{T}^\mu_{\ \ \tau} + A_\tau j^\mu$  
and $
\mc{J}^\mu= \hat{T}^\mu_{\ \ \varphi} + (A_\varphi - Q_m C)j^\mu$
appear to be gauge dependent because of the explicit presence of 
$A_\tau$ and $A_\varphi$ in them,
it is important to note that   
the vector potential $A$ that is used in these expressions is invariant under 
time translations and rotations, which is a property that is also used when
Noether's theorem is applied, and which fixes
$A_\tau$ and $A_\varphi$ uniquely 
(up to additive constants), as mentioned in section \ref{sec.knd}.
Furthermore, if $A$ is replaced by 
some gauge transformed vector potential 
$A+d\Phi$ in the Lagrangian (\ref{eq.lagr}), then 
the quantity $K^\mu$ appearing in the invariance condition (\ref{eq.sym}) also has to be modified
as $K^\mu \to K^\mu + j^\mu \partial_\tau \Phi$ 
or $K^\mu \to K^\mu + j^\mu \partial_\varphi \Phi$ (for time translations and rotations,
respectively), where $j^\mu$ is the electric current.
The effect
of this modification is that the $A_\tau$ and $A_\varphi$ appearing explicitly in the 
formulas $\mc{E}^\mu = \hat{T}^\mu_{\ \ \tau} + A_\tau j^\mu$  
and $\mc{J}^\mu= \hat{T}^\mu_{\ \ \varphi} + (A_\varphi - Q_m C)j^\mu$ remain unchanged.
Of course, the vector potential in the expressions for 
$\hat{T}^\mu_{\ \ \tau}$ and $j^\mu$ will be the gauge transformed one. 
The expressions (\ref{eq.y1}) and (\ref{eq.y2}) 
for the conserved energy and angular momentum in the case of the 
pointlike test particle can also be derived using Noether's theorem, 
and an argument analogous to the one above shows that   
the $A_\tau$ and $A_\varphi$ appearing in these expressions are also well defined.

The tensor $\hat{T}_{\mu\nu}$ coincides with the 
Einstein-Hilbert 
energy-momentum tensor \\
$-2\frac{\delta\mc{L}}{\delta g^{\mu\nu}} + g_{\mu\nu}\mc{L}$
obtained from 
the Lagrangian
(\ref{eq.lagr}), and the inequality $\hat{T}_{\mu\nu}\chi^\mu \chi^\nu \ge 0$ used above 
has the form of a null energy condition. 
One might think that the energy and angular momentum currents should be defined as 
$\mc{E}^\mu= \hat{T}^\mu_{\ \ \tau}$ and 
$\mc{J}^\mu= \hat{T}^\mu_{\ \ \varphi}$, however, these currents are not conserved, which can be seen 
by considering that their conservation 
would imply the conservation of $A_\tau j^\mu$ and $A_\varphi j^\mu$. The non-conservation of these currents 
also means that $\nabla_\mu\hat{T}^{\mu}_{\ \ \nu}\ne 0$.
By looking at the derivation of the conservation of the Einstein-Hilbert energy-momentum tensor
(see e.g.\  section E.1 of \cite{WaldGR} around equation (E.1.27)) 
it can be seen that
the obstacle to the conservation of $\hat{T}_{\mu\nu}$ is 
the presence of the fixed electromagnetic field.

\subsection{Scalar and electromagnetic test fields}
\label{sec.ft2}

The Lagrangian density of the scalar and electromagnetic fields in Kerr-Newman spacetime is
\beq
\mc{L}=-g^{\mu\nu}(\partial_\mu-ie\tilde{A}_\mu)\psi^*(\partial_\nu+ie\tilde{A}_\nu)\psi-m^2\psi^*\psi-\frac{1}{16\pi}\tilde{F}_{\mu\nu}\tilde{F}^{\mu\nu},
\eeq
where $\tilde{F}_{\mu\nu}=\partial_\mu \tilde{A}_\nu- \partial_\nu \tilde{A}_\mu$ and the tilde is used to 
distinguish the vector potential of the full electromagnetic field from the vector potential of the 
electromagnetic field of the dyonic Kerr-Newman black hole introduced in section \ref{sec.knd}. 
The electric current is 
\beq
j^\mu=ie[\psi^*(\partial^\mu+ie\tilde{A}^\mu)\psi-\psi(\partial^\mu-ie\tilde{A}^\mu)\psi^*].
\eeq
In the present setting the Einstein-Hilbert energy-momentum tensor 
\begin{eqnarray}
T_{\mu\nu} & = & -2\frac{\delta\mc{L}}{\delta g^{\mu\nu}} + g_{\mu\nu}\mc{L} \nonumber\\
& = & (\partial_\mu-ie\tilde{A}_\mu)\psi^* (\partial_\nu +ie\tilde{A}_\nu)\psi
+ (\partial_\mu+ie\tilde{A}_\mu)\psi (\partial_\nu -ie\tilde{A}_\nu)\psi^* \nonumber \\
\label{eq.EH}
&& +\frac{1}{4\pi} \tilde{F}_{\mu\lambda}{\tilde{F}_\nu}^{\ \lambda}
+ g_{\mu\nu}\mc{L}
\end{eqnarray}
is conserved (i.e.\ $\nabla_\mu {T^\mu}_\nu =0$) and is suitable for defining the energy and angular momentum 
currents as 
\beq
\mc{E}^\mu={T^\mu}_\tau,\qquad \mc{J}^\mu={T^\mu}_\varphi.
\eeq 
These currents are conserved 
(i.e.\ $\nabla_\mu\mc{E}^\mu=0$ and $\nabla_\mu\mc{J}^\mu=0$)
because
$\partial/\partial\tau$ and $\partial/\partial\varphi$ are Killing vectors and 
$\nabla_\mu {T^\mu}_\nu =0$. $\mc{E}^\mu$ and $\mc{J}^\mu$ are also clearly gauge invariant. 
The same definition is taken for the energy and angular momentum 
currents
in \cite{Semiz1}.

The charge, energy and angular momentum fluxes through the event horizon are given by 
\begin{eqnarray}
\label{eq.dqdt2}
\frac{dQ}{d\tau} & = & - \int_H \sqrt{-g}\ j^r \, d\theta d \varphi\\
\label{eq.dedt2}
\frac{dE}{d\tau} & = & \phantom{-}\int_H \sqrt{-g}\ {T^r}_\tau \, d\theta d \varphi\\
\label{eq.dldt2}
\frac{dL}{d\tau} & = & - \int_H \sqrt{-g}\ {T^r}_\varphi \, d\theta d \varphi\ .
\end{eqnarray}
We assume 
that the field $\psi$ goes to zero as $\tau\to\infty$,
in accordance with the fundamental assumption 
that the final state of the physical process under consideration
is a dyonic Kerr-Newman state.
The vector potential, on the other hand, will become $A+dQ_e A_e$ 
(up to gauge transformation)
as $\tau\to\infty$ due to the change $dQ_e$ of the charge of the black hole. This change of the 
electromagnetic field implies that the energy and the angular momentum of the electromagnetic field around the 
black hole also changes, which has to be 
taken into account in the calculation of $dM$ and $dJ$.   
Thus $dM$ and $dJ$ are given by 
\begin{eqnarray}
dM & = & \int_{-\infty}^\infty  \frac{dE}{d\tau} d\tau 
-
\int_{r_+}^{\infty} dr \int \sqrt{-g}\ \left.{T^\tau}_\tau \right|_{\tilde{A}=A+dQ_e A_e,\, \psi=0}\, 
d\theta d \varphi \nonumber\\
\label{eq.dm2}
&& 
+
\int_{r_+}^{\infty} dr \int \sqrt{-g}\, \left.{T^\tau}_\tau \right|_{\tilde{A}=A,\, \psi=0} \,  
d\theta d \varphi
\end{eqnarray}
and
\begin{eqnarray}
dJ & = & \int_{-\infty}^\infty  \frac{dL}{d\tau} d\tau 
+
\int_{r_+}^{\infty} dr \int \sqrt{-g}\ \left.{T^\tau}_\varphi \right|_{\tilde{A}=A+dQ_e A_e,\, \psi=0}\, 
d\theta d \varphi
 \nonumber\\
\label{eq.dj2}
&& -
\int_{r_+}^{\infty} dr \int \sqrt{-g}\, \left.{T^\tau}_\varphi \right|_{\tilde{A}=A,\, \psi=0} \,  
d\theta d \varphi \ .
\end{eqnarray}
The second term on the right hand side of (\ref{eq.dm2}) and (\ref{eq.dj2})
gives the energy and angular momentum, respectively,
of the electromagnetic field around the black hole at $\tau\to \infty$, whereas the 
third terms give the energy and angular momentum
of the electromagnetic field around the black hole in the initial state.
$dQ_e$ is given by 
$dQ_e=\int_{-\infty}^\infty  \frac{dQ}{d\tau} d\tau$, as in section \ref{sec.ft1}.
We note that in \cite{Semiz1} the change of 
the energy and angular momentum of the electromagnetic field around the 
black hole is included in $dM$ and $dJ$ by 
considering fluxes through spherical surfaces of radius $r\to\infty$ rather than $r=r_+$.

Aiming to derive an equation similar to (\ref{eq.oo}), we consider now the quantity $dM-\Omega_H dJ$.
From (\ref{eq.dedt2}),  (\ref{eq.dldt2}) and (\ref{eq.chi}) it is easy to see that 
the contribution of the first terms on the right hand side of (\ref{eq.dm2}) and (\ref{eq.dj2})
to $dM-\Omega_H dJ$ is 
$\int_{-\infty}^\infty d\tau \int_H \sqrt{-g}\ T_{\mu\nu}\omega^\mu \chi^\nu \, d\theta d\varphi$.
Since $A_e$ and $A_m$ are known explicitly, the contribution of the second and third terms 
on the right hand side of (\ref{eq.dm2}) and (\ref{eq.dj2}) can also be evaluated. This task can be 
simplified by partial integrations and by using the properties of $A_e$, $A_m$ and of the corresponding  
electromagnetic fields. In addition, those terms that are higher than first order in $dQ_e$ should be neglected. 
One finds that all integrals can be evaluated trivially except for one integral
over $\theta$,
and the final result is that the contribution of the terms in question is $\Phi_H dQ_e$. 
Thus
\beq
\label{eq.oo2}
\int_{-\infty}^\infty d\tau \int_H \sqrt{-g}\ T_{\mu\nu}\omega^\mu \chi^\nu   \, d\theta d \varphi =
dM-\Omega_H dJ -  \Phi_H dQ_e\ ,
\eeq
which is analogous to (\ref{eq.oo}) in section \ref{sec.ft1}. 
From the fact that $\chi^\mu$ is a null vector at the event horizon and from the form
(\ref{eq.EH}) of $T_{\mu\nu}$ it is obvious that at the event horizon $T_{\mu\nu}$ satisfies 
the inequality $T_{\mu\nu}\chi^\mu \chi^\nu\ge 0$ ---a null energy condition---for arbitrary $A_\mu$ and $\psi$ functions. 
Taking into consideration (\ref{eq.rho}), this implies that  $T_{\mu\nu}\omega^\mu \chi^\nu\ge 0$ also
holds for the integrand on the left hand side of (\ref{eq.oo2}), hence 
\beq
\label{eq.yy}
dM-\Omega_H dJ -  \Phi_H dQ_e\ge 0. 
\eeq 
This inequality, which has the same form as (\ref{eq.y}), 
implies $d\eta\ge 0$ in the case when the black hole is extremal, 
indicating that the cosmic censorship is not violated.

We note that Noether's theorem gives the conserved energy and angular momentum currents
\begin{eqnarray}
\label{eq.nc1}
\mc{E}^\mu & = & {T^\mu}_\tau - \frac{1}{4\pi\sqrt{-g}}\partial_\rho(\sqrt{-g}\tilde{A}_\tau \tilde{F}^{\rho\mu})\\
\label{eq.nc2}
\mc{J}^\mu & = & {T^\mu}_\varphi - \frac{1}{4\pi\sqrt{-g}}\partial_\rho(\sqrt{-g}(\tilde{A}_\varphi-Q_mC) \tilde{F}^{\rho\mu}).
\end{eqnarray}
The additional terms 
$ \frac{-1}{4\pi\sqrt{-g}}\partial_\rho(\sqrt{-g}\tilde{A}_\tau \tilde{F}^{\rho\mu})$ and
$ \frac{-1}{4\pi\sqrt{-g}}\partial_\rho(\sqrt{-g}(\tilde{A}_\varphi-Q_mC) \tilde{F}^{\rho\mu})$
are of the form $\nabla_\nu f^{\mu\nu}$, where $f^{\mu\nu}$ is antisymmetric. Currents of this form are automatically conserved regardless of the value of $f^{\mu\nu}$. 
It is not difficult to verify using partial integration that these terms 
do not give any contribution to $dM$ and $dJ$, therefore 
the definitions (\ref{eq.nc1}) and (\ref{eq.nc2}) also lead to the 
results (\ref{eq.oo2}) and (\ref{eq.yy}).

We also note finally that the presence of the scalar field is not essential in the derivation above.
If it is omitted, then the case of a purely electromagnetic test field is obtained.



\section*{Appendix. Noether's theorem}
\label{sec.noether}

\renewcommand{\theequation}{A.\arabic{equation}} 
\setcounter{equation}{0}

Let the action of a physical system described by a collection of real fields $\Phi_i(x^a)$ on an $n$-dimensional spacetime be
\beq
\mc{S}=\int dx^1 dx^2 \dots dx^n\, \mb{L}(\Phi_i(x^a),\partial_b\Phi_i(x^a),x^a),
\eeq
with Lagrangian  $\mb{L}(\Phi_i(x^a),\partial_b\Phi_i(x^a),x^a)$.
The equations of motions are the Euler-Lagrange equations
\beq
\label{eq.el}
\frac{\partial \mb{L}}{\partial \Phi_i}-D_\mu\frac{\partial \mb{L}}{\partial(\partial_\mu\Phi_i)}=0.
\eeq
The notation $D_\mu$ is used for the total derivative with respect to $x^\mu$.
If, for example, $f$ is a function of $x^a$, then $D_\mu f=\partial_\mu f $, whereas for a function $f(\Phi_i, x^a)$ we have
$D_\mu f=\frac{\partial f}{\partial \Phi_i} \partial_\mu \Phi_i + \frac{\partial f}{\partial x^\mu}$.  

Assume that $\Phi_i$ satisfy the Euler-Lagrange equations, and the invariance condition  
\beq
\label{eq.sym}
\frac{\partial \mb{L}}{\partial\Phi_i}\Delta\Phi_i+\frac{\partial \mb{L}}{\partial(\partial_\mu\Phi_i)}D_\mu (\Delta\Phi_i)= D_\mu K^\mu
\eeq
holds with some functions $\Delta\Phi_i$ and $K^\mu$. $\Delta\Phi_i$ denote the change of the fields under an infinitesimal transformation 
$\Phi_i\to \Phi_i+\epsilon\Delta\Phi_i$. The expression on the left hand side is the change of $\mb{L}$ under this transformation.  
Now it is straightforward to see, using (\ref{eq.el}) and (\ref{eq.sym}), that the current 
\beq
\label{eq.curr}
j^\mu=\frac{\partial \mb{L}}{\partial(\partial_\mu\Phi_i)}\Delta\Phi_i-K^\mu
\eeq
is conserved, i.e.\ 
\beq
D_\mu j^\mu=0.
\eeq
This theorem is independent of any metric structure on the spacetime manifold.

In section \ref{sec.ft} we have $\mb{L}=\sqrt{-g} \mc{L}$; for time translations 
\beq
\Delta \psi=-\partial_\tau \psi,\quad  \Delta \psi^*=-\partial_\tau \psi^*,\quad
\Delta\tilde{A}_\mu=-\partial_\tau \tilde{A}_\mu,\quad
K^\mu= -{\delta^\mu}_\tau \sqrt{-g} \mc{L}\,;
\eeq
for rotations
\beq
\Delta \psi=-\partial_\varphi \psi,\quad
\Delta \psi^*=-\partial_\varphi \psi^*,\quad 
\Delta\tilde{A}_\mu=-\partial_\varphi \tilde{A}_\mu,\quad
K^\mu= -{\delta^\mu}_\varphi \sqrt{-g} \mc{L}\,.
\eeq
For global $U(1)$ gauge transformations we have 
\beq
\Delta \psi=i \psi,\quad \Delta \psi^*=-i \psi^*,\quad
K^\mu=0.
\eeq
The invariance condition is satisfied for any fields in these cases, 
not only for the solutions of the Euler-Lagrange equations.

\section*{Acknowledgment}

The author would like to thank Istv\'an R\'acz for useful discussions.

\end{document}